\documentclass[aps,prl,twocolumn]{revtex4}

\usepackage{graphicx}

\usepackage{color}
\usepackage[normalem]{ulem}

\begin{document}

%\title{Droplet formation in \out{trapped} \new{quantum} dipolar
%\out{systems of} \new{bosons}}
\title{Droplets of trapped quantum dipolar bosons}

\author{A. Macia, J. S\'anchez-Baena, J. Boronat, and F. Mazzanti}

% \affiliation{$^1$ Arham asylum, Gotham city}

\affiliation{Departament de F\'{\i}sica, Campus Nord
  B4-B5, Universitat Polit\`ecnica de Catalunya, E-08034 Barcelona, Spain}

%\date{\today}

\begin{abstract}

Strongly interacting systems of dipolar bosons in three dimensions
confined by harmonic traps are analyzed using the exact Path Integral
Ground State Monte Carlo method. By adding a repulsive two-body
potential, we find a narrow window of interaction parameters leading
to stable ground-state configurations of droplets in a crystalline
arrangement. We find that this effect is entirely due to the
interaction present in the Hamiltonian without resorting to additional
stabilizing mechanisms or specific three-body forces.  We analyze the
number of droplets formed in terms of the Hamiltonian parameters,
relate them to the corresponding $s$-wave scattering length, and
discuss a simple scaling model for the density profiles.  Our results
are in qualitative agreement with recent experiments showing a quantum
Rosensweig instability in trapped Dy atoms.

\end{abstract}


\maketitle

Dipolar effects in quantum gases have been considered of major
experimental and theoretical interest in the last decade since the
initial studies of dilute clouds of Cr atoms, which present a
relatively large magnetic dipolar moment.  In the pioneering
experiments of Ref.~\cite{Griesmaier2005}, the two-body scattering
length of a cloud of $^{52}$Cr atoms was drastically reduced by
bringing it close to a Feshbach resonance.  In this way, dipolar
effects were enhanced and interesting new features, not observed
before in other species like Rb or Cs, appeared. The long range and
anisotropic character of the dipolar interaction has been largely
explored since then, leading to interesting new phenomena such as
$d$-wave superfluidity or $d$-wave collapse~\cite{Lahaye_2007,
  Lahaye_2008}.  All these experiments have opened new perspectives on
the field of dipolar quantum physics, and new systems with stronger
dipolar interactions have since then been explored. The most promising
ones, consisting initially in ultracold polar molecules of K and Rb or
Cs and Rb, are unfortunately problematic due to the inherent
difficulty to bring them down to the quantum degeneracy limit,
although recent progress have been achieved with NaK~\cite{park} and
NaRb~\cite{guo} molecules.  Alternatively, Bose-Einstein condensate
(BEC) states of Er~\cite{aikawa_12} and Dy~\cite{lu} have recently
been produced, enabling for instance to observe the deformation of the
Fermi surface~\cite{Aikawa_2014}, or the influence of the anisotropy
on the superfluid to Mott insulator transition in the extended Hubbard
model made with dipoles~\cite{Baier_2016}.

The anisotropy of the dipolar interaction plays a fundamental role on
the behavior of the system, with different regimes and phases
depending on the geometry and dimensionality.  The particular form of
the dipole-dipole potential makes the interaction be attractive or
repulsive depending on the relative orientation of the dipoles,
according to the expression
\begin{equation}
  V_{dd}({\bf r}) = {C_{dd}\over 4\pi} \left[ { {\bf \hat p}_1\!\cdot\!{\bf \hat
      p}_2 - 3 ({\bf \hat p}_1\!\cdot\!{\bf \hat r}) ({\bf \hat
      p}_2\!\cdot\!{\bf \hat r}) \over r^3 }
    \right] \ ,
  \label{dipdippot_b1}
\end{equation}
where $C_{dd}$ sets the strength of the interaction that is
proportional to the square of the (magnetic or electric) dipolar
moment, ${\bf p}_j$ is the dipolar moment itself, and ${\bf r}$ is the
relative position vector of the two interacting dipoles.  The
particular form of this interaction leads to surprising new features
not present in other systems, like stripe phases in two-dimensional
(2D) Bose systems~\cite{Macia_2012}. Similar phases in Fermi systems
have also been predicted~\cite{Yamaguchi_2010, Parish_2012}, although
these are more controversial~\cite{Matveeva_2012}.

One of the most interesting phenomenon recently reported in the field
of dipolar quantum gases is the formation of self-bound droplets when
a gas of trapped $^{164}$Dy atoms is brought to the regime of mean
field collapse~\cite{Kadau_2016,Ferrier-Barbut_2016}. More
interestingly, the resulting set of droplets are reported to arrange
themselves in a crystalline structure that becomes more clearly
visible when the number of droplets increases.  This phenomenon
resembles the Rosensweig instability of a classical ferrofluid
appearing when the magnetization increases, and thus the dipolar gas
can be considered as a first realization of a quantum ferrofluid.
From the theoretical side, this is a clear example of a situation
where beyond mean-field effects drastically determine the physics of
the system. Initial mechanisms based on the inclusion of three-body
forces were reported to produce both effects~\cite{Bisset_2015,
  Blakie_2016, Kui_Tian_2016} (droplet formation and crystallization),
although this mechanism does not seem to be fully compatible with
experimental data~\cite{Ferrier-Barbut_2016}.  Beyond mean field
effects at the level of the Lee-Huang-Yang correction
(LHY)~\cite{Lee_1957}, as proposed
in~\cite{Wachtler_2016,Wachtler_2016_B,Bisset_2016}, produce
equivalent effects.
%% and are now understood to provide a more plausible theoretical
%% model.
%% This is mostly due to the fact that, at the mean field level,
%% the contribution of the two-body contact force and the dipolar term
%% %% \new{(through the LHY correction) }
%% seem to be similar but of different
%% sign, so that they essentially cancel.
A similar stabilization
mechanism has been recently proposed to make possible the formation of
liquid droplets in attractive Bose-Bose mixtures~\cite{petrov}.
%% An important question here is to \new{determine the range of
%%   applicability of the LHY correction since the droplets could become
%%   quite dense and it is known that this term cannot account for more
%%   strongly correlated Bose gases.  }

%% \out{decide whether this
%%   correction is strong enough to lead to the crystallization of the
%%   resulting droplets, since previous analysis of LHY corrections in
%%   model systems do not lead to phase transitions of this kind in a
%%   weakly interacting gas of bosons}~\cite{Giorgini_1999,
%%   Mazzanti_2003}.

In this work, we address the problem of droplet formation of trapped
dipolar bosons from a microscopic point of view, using the stochastic
Path Integral Ground State (PIGS) method~\cite{Sarsa_2000, Galli_2003,
  Rota_2010}. Starting from a variational {\em ansatz} for the
ground-state wave function, propagation in imaginary time leads to a
statistically exact representation of the actual ground state of the
system that is used to sample relevant observables. Differently from
the perturbative approaches discussed before, PIGS includes
correlations induced by the interactions to all levels.  Accurate
fourth-order approximations of the many-body propagator are employed
in order to guarantee convergence with the lowest possible number of
imaginary time steps. In particular, we have employed the 4A
fourth-order short time expansion propagator proposed by Chin and Chen
in Ref.~\cite{Chin_2002}. Recently, Saito~\cite{Saito_2016} performed
a path integral Monte Carlo simulation of a single droplet but the
formation of the ordered array of droplets found in experiments was
not reproduced.

In the following we describe a system of $N$ trapped dipolar bosons
of mass $m$ by the Hamiltonian
\begin{equation}
  H = -{\hbar^2 \over 2m} \sum_j {\nabla_j^2} +
  \sum_j V_{ho}({\bf r}_j) + \sum_{i<j} V_\sigma(r_{ij}) +
  \sum_{i<j} V_{dd}({\bf r}_{ij}) \ ,
  \label{Hamiltonian}
\end{equation}
where
\begin{equation}
  V_{ho}({\bf r}) =
  {1\over 2}\,m\left(\omega_x^2 x^2 +
  \omega_y^2 y^2 + \omega_z^2 z^2 \right)
  \label{V_HO}
\end{equation}
is the three-dimensional harmonic trapping potential of frequencies
$\omega_x, \omega_y $ and $\omega_z$, and $V_\sigma(r)$ is a
short-range two-body repulsive interaction of the form
$V_\sigma(r)=(\sigma/r)^{12}$, where the parameter $\sigma$ can be
varied to tune the $s$-wave scattering length.  This interaction is so
steep at short distances that in practical terms it is hardly
distinguishable from a pure hard core. In the following we
use dimensionless quantities, introducing a characteristic
length scale $r_0=m C_{dd}/(4\pi\hbar^2)$ and a characteristic energy
scale $E_0=\hbar^2/(m r_0^2)$.

Inspired by recent experimental measurements~\cite{Kadau_2016}, we
have considered different pancake-shaped harmonic traps with
oscillator lengths $a_x=a_y > a_z$ ($a_\alpha=\sqrt{\hbar/(m
  \omega_\alpha)}$).  Two different choices, leading to more
than one stable droplet, have been used: Trap1 with $a_x=1.20,
a_z=0.6$, and Trap2 with $a_x=1.38, a_z=0.45$. Additionally, a Trap3
model yielding a single stable droplet with $a_x=1.73$ and $a_z=1.00$
has also been considered.
%% {\bf PARA QU\'E VALORES DE $\sigma$ ??}{\bf ES CIERTO QUE
%%   TODASLAS FIGURAS DE GOTAS CORRESPONDEN A UN MISMO VALOR DE $\sigma$,
%%   NO? LO DIGO PQ EN LOS EXPERIMENTOS USAN UN VALOR DEL SCATT LENGTH
%%   FIJO, QUE CORRESPONDE A UNA $\sigma$ CONCRETA}
Notice that the first choice is close to the experimental setting of
Ref.~\cite{Kadau_2016} where $a_x\sim a_z\sqrt{3}$, while in the
second model the aspect ratio is different. With these settings,
different values of $\sigma$ lead to different ground-state
configurations. In general, the resulting ground state is a gas, but
we have checked that for the number of particles in the few-hundreds
range used, there is a narrow window of values, spanning the range
$\sigma \in[0.24,0.28]$, where stable configurations of droplets are
obtained.
%% {\bf ES EL MISMO RANGO PARA TODAS
%%   LAS SIMULACIONES QUE HEMOS HECHO, NO?  VARIANDO EL NUMERO DE
%%   PARTICULAS, ETC...}

\begin{figure}[t]
\begin{center}
\includegraphics*[width=0.40\textwidth]{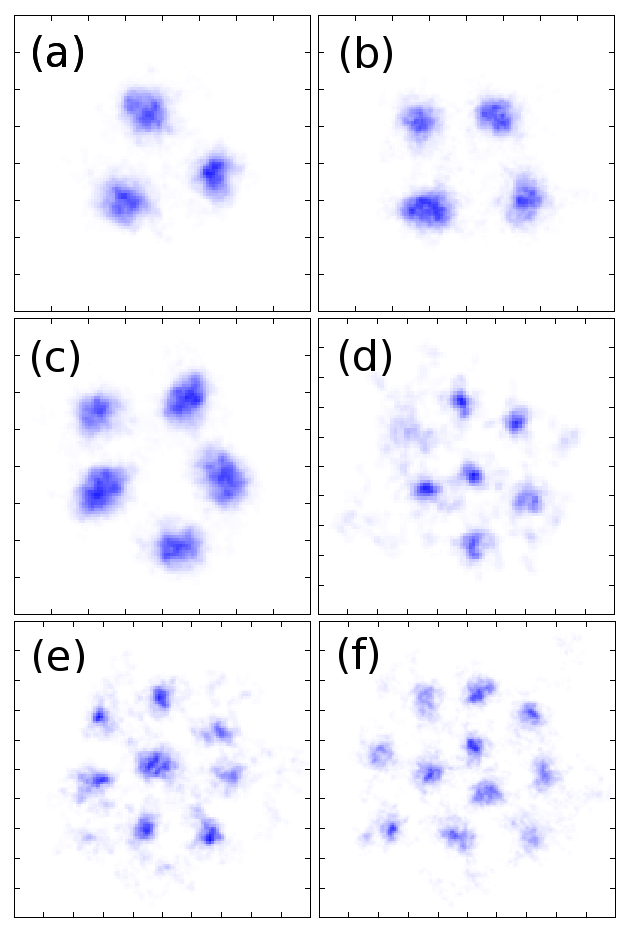}
\end{center}
\caption{Spatial density distributions of the stable droplet
  configurations obtained from the quantum simulations. Axes in (a), (b) and
  (c) range from $-4 r_0$ to $4 r_0$ while those in (d), (e) and (f)
  span the range from $-5 r_0$ to $5 r_0$.}
\label{fig-droplets}
\end{figure}

\begin{figure}[h!]
\begin{center}
  \includegraphics*[width=0.45\textwidth]{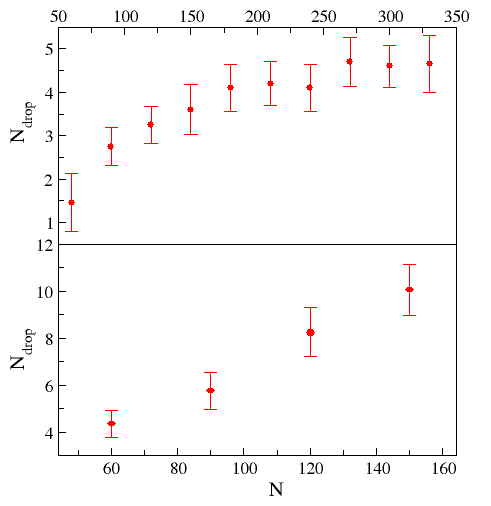}
\end{center}
\caption{Number of droplets as a function of the total number of
  particles in the simulation. Upper and lower plots correspond to the
  Trap1 and Trap2 set of parameters explained in the text,
  respectively.}
\label{fig_NumDrop_vs_N}
\end{figure}

Results of the ground state configurations obtained from the PIGS
simulations are reported in Fig.~\ref{fig-droplets}. Plots (a), (b)
and (c) correspond to Trap1, and (d), (e) and (f) to Trap2. Each plot
corresponds to $\sigma=0.28$ but different number of particles,
$N=120, 150, 270, 90, 120, 150$ (from (a) to (f)), leading to an
increasing number of droplets as can be seen in the figure.  These
representative configurations show well-formed droplets that are
easily distinguishable from each other, although not every
configuration after thermalization is that neat. This indicates that
at some point the system may reach metastable configurations.  Notice
that, despite the number of particles in all cases is similar, Trap1
tends to form fewer but larger droplets than Trap2. This fact is more
clearly seen in Fig.~\ref{fig_NumDrop_vs_N}, where the average number
of droplets as a function of the number of particles for the Trap1 and
Trap2 models is reported. The error bars indicate the fluctuations
that results from the averaging procedure performed over the different
configurations used in each case. The amplitude of the error bars can
then be taken as a measure of the degree of metastability of the
analyzed states.

As it can be seen, in the Trap2 case the observed dependence is
linear, in agreement with the behavior observed in the experiments of
Ref.~\cite{Kadau_2016}. However, in the Trap1 case an approximately
linear dependence at the beginning of the curve is followed by a
change of tendency, where one can not decide if the average number of
droplets saturates or increases very slowly with the total number of
particles in the simulation. A similar behavior has been reported in
Ref.~\cite{Wachtler_2016} using the LHY correction, 
while in that case the number of atoms in the system is much larger.

%% \vspace*{3mm}
%% \hrule
%% Discusion of the density profiles
%% \hrule
%% \vspace*{3mm}

The stochastic sampling of the different configurations allows for the
statistical analysis of the density profiles of the droplets formed in
each case.  In general, changing the model parameters ($\sigma$,
number of particles and trapping frequencies) leads to different
droplet configurations and density profiles. We analyze the specific
cases where a reduced number of droplets appear, due to the
restrictions on the total number of particles employed. We report
results for the normalized and marginalized $n(x), n(y)$ and $n(z)$
density profiles. The normalization has been set such that the
integral of each separate profile is one, i.e., $\int
d\alpha\,n(\alpha)=1$ for $\alpha=\{x,y,z\}$. Notice that due to axial
symmetry, the $n(x)$ and $n(y)$ profiles are identical up to
statistical errors.

\begin{figure}[t]
\begin{center}
  \includegraphics*[width=0.40\textwidth]{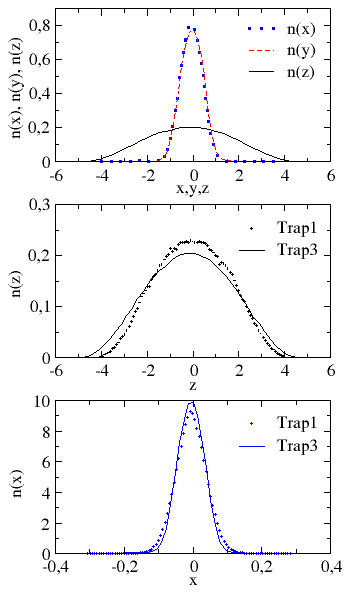}
\end{center}
\caption{Upper panel: normalized marginal density profiles $n(x)$,
  $n(y)$ and $n(z)$ of the single droplet obtained in the Trap3 setup
  with $N = 150$.  Middle and lower panels: scaled $n(z)$ and $n(x)$
  density profiles for the droplets found in the Trap1 and Trap3
  setups, with $120$ and $150$ particles per droplet, respectively.}
\label{fig_dens_prof}
\end{figure}

The obtained density profiles are shown in
Fig.~\ref{fig_dens_prof}. The upper plot shows $n(x)$, $n(y)$ and
$n(z)$ for the Trap3 model with $N=150$ particles. All three profiles
are approximately Gaussian but with marked differences in width when
the $z$ component is compared to the other two.  This is due to the
combined effect of the anisotropy of the dipolar interaction and the
shape of the confining trap.  Despite the fact that the trap is
tighter in the $z$ direction, the droplet is larger along the $z$
axis.  In fact, the dipolar interaction is attractive along this line,
and thus particles prefer to arrange themselves in head-to-tail
configurations with a minimum distance imposed by the core of
$V_\sigma(r)$. Therefore, the \textit{height} of the droplet is mainly
constrained by $\omega_z$.  However, the width $r_{x}\simeq r_{y}$
along the radial directions is set by a more complex combination of
parameters, as can be seen from the fact that a single droplet does
not cover the whole available space.  Actually, according to the
obtained profiles, the self-induced confinement along the radial
direction is much stronger, forming prolate stable droplets.

One simple model that captures these features assumes that the average
density of the droplet does not change much in the range of parameters
analyzed in this work. Assuming also that the height of the droplet is
essentially fixed by $\omega_z$, a simple 1D harmonic oscillator model
leads to a height directly proportional to the $a_z$ oscillator
length. These two assumptions also imply a simple dependence of
$r_{x}$ ($\sim r_y$) on $\sqrt{N_d/a_z}$, with $N_d$ the total number
of particles in the droplet. In order to check the accuracy of this
simple model, the middle and lower panels in Fig.~\ref{fig_dens_prof}
depict scaled $z$ and $x$ profiles for the Trap1 and Trap3 cases and a
total number of particles $N=120$ and $N=150$, respectively. As it can
be seen, this simple model seems to capture the main trends reasonably
well, although the fine detail due to more complex contributions is
missing.

\begin{figure}[t]
\begin{center}
\includegraphics*[width=0.40\textwidth]{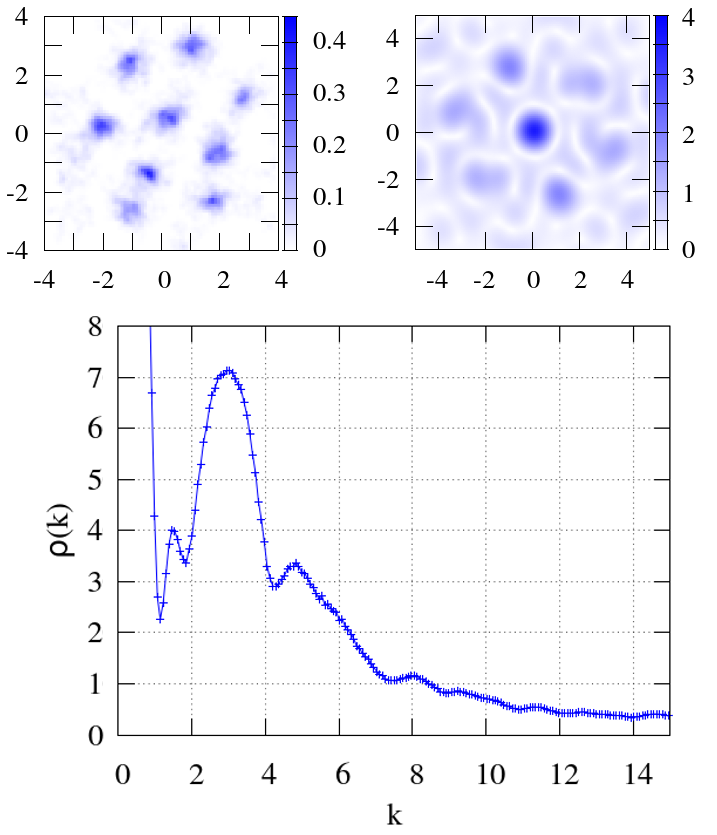}
\end{center}
\caption{Upper left panel: spatial density distribution of the
  averaged stable Trap2 configuration with nine droplets. Upper right
  panel: density distribution in momentum space of the configuration
  shown in the upper left panel. Lower panel: angularly averaged
  density in momentum space.}
\label{fig_rhok}
\end{figure}

Following the analysis of Ref.~\cite{Kadau_2016}, we
report in Fig.~\ref{fig_rhok} the Fourier spectrum of the averaged stable
Trap2 configuration with nine droplets shown in the upper left
panel. The resulting Fourier transform density is depicted in the
upper right panel. Notice that the latter shows clear regularities that
are due to the crystalline spatial distribution of the droplet
configuration. The lower panel shows the radially averaged density
$\rho\left(\sqrt{k_x^2+k_y^2}\right)$ in momentum space. This quantity
presents a large value around the origin proportional to the average
number of particles and, more importantly, a clear peak around 
$k\sim3$ which reflects
the spatial periodicity of the underlying triangular lattice
formed by the droplets.

%% \vspace*{3mm}
%% \hrule
%% Discusion scattering length
%% \hrule
%% \vspace*{3mm}

A better comparison with experiments requires the analysis of the 
range of scattering lengths
covered in our simulations. As previously mentioned, the interaction
parameters where stable configurations of droplets are found span the
range from $\sigma=0.24$ to $0.28$ in dipolar units.  Relating these
to the $s$-wave scattering length $a_0$ is not 
immediately trivial, as it is known that the
presence of the dipolar interaction in the trap modifies its
value~\cite{Bortolotti_2006,Ronen_2006}.  Following a standard
procedure, we have determined the low-energy behavior of the
coupled-channel T-matrix associated to the two-body potential
employed, including the full dipole-dipole interaction. 
%% The latter mixes states of different angular momentum due to its anisotropic
%% character. As we are dealing with a three-dimensional problem, the
%% long-range behavior of the dipole-dipole interaction leads to non-zero
%% contribution from several scattering lengths of different
%% channels. However, only the $s$-wave scattering length $a_0$ changes
%% with $\sigma$.
Table~\ref{table_scattlength} reports the values of
$a_0$ obtained for the different $\sigma$'s used in this work.  In
contrast to the experiments~\cite{Kadau_2016, Ferrier-Barbut_2016},
in our case all situations leading to stable droplet configurations
correspond to negative values of $a_0$, appearing before a first
resonance is found. This is due to the attractive components of the
dipolar interaction.
%% In general, a negative scattering length in a
%% dilute Bose gas leads to collapse, although in our case the
%% anisotropy of the interaction, together with the strong hard core
%% repulsion, originates a narrow window of negative scattering length
%% values where an array of self-bound droplets is formed.
In any case, the
range of $\sigma$'s (and therefore scattering lengths) leading to
droplet formation may change when a much larger number of particles
and/or different trapping frequencies are used, as in the
experiments. The dependence of the number of droplets on
$a_0$ is also indicated in the third
and fourth columns of Table~\ref{table_scattlength} for the Trap1 set
of parameters. As it can be seen, increasing $a_0$ leads to a larger
number of smaller droplets.
% \out{ This is a general tendency that has also
% been observed in the different simulations performed.}

\begin{table}[t]
\begin{center}
{
\tabcolsep=4mm
\def\arraystretch{1.3}
\begin{tabular}{|c|ccc|} \hline
  $\sigma$ & $a_0$ & $\langle N_{\text{droplet}} \rangle$ &
  $\Delta\langle N_{\text{droplet}} \rangle$ \\ \hline
  0.24  & -1.417   &    1.60   &  0.44  \\
  0.25  & -0.868   &    1.85   &  0.44  \\
  0.26  & -0.618   &    2.79   &  0.31  \\
  0.27  & -0.472   &    2.85   &  0.69  \\
  0.28  & -0.374   &    3.25   &  0.19  \\ \hline
\end{tabular}
\caption{$s$-wave scattering length, average number of droplets
  $\langle N_{\rm drop}\rangle$ and standard deviation
  $\Delta\langle N_{\rm drop}\rangle$ for the Trap1 model with $N=120$
  particles, as a function of $\sigma$.
}
\label{table_scattlength}
}
\end{center}
\end{table}

In summary, we have used the Path Integral Ground State algorithm with
a fourth-order propagator to determine the ground state structure of a
system of trapped dipolar bosons with an additional repulsive
$(\sigma/r)^{12}$ core. 
Contrarily to previous perturbative estimates, 
PIGS is exact and relies only on the Hamiltonian of the system
and the chosen geometry. The formation of self-bound droplets and its
arrangement in a crystal lattice is on the basis of the microscopic
Hamiltonian and,
therefore, there is no need to resort to three-body interactions
%% to stabilize the mean-field.
%% thus, the introduction of ad-hoc three-body interactions to stabilize
%% the mean-field system is just an artifact.
We find that, for a few hundred dipoles, there is a window of
$\sigma$'s corresponding to negative $s$-wave scattering lengths where
the ground state forms a crystal of stable droplets, rather than
collapsing to a single one or remaining as a gas. We have analyzed the
density profiles of the droplets to find that a very simple model is
able to qualitatively explain how these scale when the number of
particles and trapping parameters are changed. The droplets are
prolate in the $z$ direction with a height which increases with $a_z$,
resembling the formation of filaments or stripes already predicted to
exist in quantum tilted dipoles in two
dimensions~\cite{stripes}. Finally, the analysis of the density
profiles in momentum space supports the general picture of crystalline
order in the droplet configurations. Work is in progress now to
determine the superfluidity of the system and its dependence on the
temperature using the path integral Monte Carlo method. A supersolid
scenario seems plausible at very low temperature.

%% \vspace*{5mm}
%% \hrule\hrule\hrule
%% \vspace*{5mm}

\vfill

\begin{acknowledgments}
We acknowledge partial financial support from the MICINN (Spain) Grant 
No.~FIS2014-56257-C2-1-P.

\end{acknowledgments}

 \end{document}